 \documentclass[final,5p,times,twocolumn,authoryear]{elsarticle}

\usepackage{amsmath}
\usepackage{amssymb}
\usepackage{lipsum}
\usepackage{siunitx}
\usepackage{url}
\usepackage{hyperref}

\journal{Astronomy $\&$ Computing}

\begin{document}

\begin{frontmatter}



\title{Extending Galactic Foreground Emission 
with Neural Networks}

 
\author[1,2,3]{Giuseppe  Puglisi}
\affiliation[1]{organization={Dipartimento di Fisica e Astronomia, Universit\`a degli Studi di Catania},
            addressline={via S. Sofia, 64 }, 
            city={Catania},
            postcode={95123}, 
            state={Italy},
            country={}}
\affiliation[2]{organization={INFN - Sezione di Catania},
            addressline={via S. Sofia, 64 }, 
            city={Catania},
            postcode={95123}, 
            state={Italy},
            country={}}

\affiliation[3]{organization={INAF - Osservatorio Astrofisico di Catania},
            addressline={via S. Sofia 78}, 
            city={Catania},
            postcode={95123}, 
            state={Italy},
            country={}}

\author[4,5]{Avinash Anand } 
\author[4,5]{Marina Migliaccio}
\affiliation[4]{organization={Dipartimento di Fisica, Università di Roma Tor Vergata},
            addressline={Via della Ricerca Scientifica, 1 }, 
            city={Roma},
            postcode={00133}, 
            state={Italy},
            country={}}
\affiliation[5]{organization={INFN Sezione di Roma2, Università di Roma Tor Vergata},
            addressline={Via della Ricerca Scientifica, 1 }, 
            city={Roma},
            postcode={00133}, 
            state={Italy},
            country={}}

\begin{abstract}
We introduce an innovative approach employing Cycle Generative Adversarial Networks (Cycle-GANs) to accurately simulate Carbon Monoxide (CO) emissions by learning features identified in thermal dust emission maps from the \emph{Planck} satellite alongside HI data from   {HI4PI survey}. Our training dataset is complemented by the targets represented by the two rotational transition lines of CO ($J:1-0,\, 2-1$) provided by the \emph{Planck} satellite. We ensure the robustness of our dataset by focusing on  regions with a signal-to-noise ratio (SNR) exceeding 8. The outcomes, assessed utilizing angular power spectra and Minkowski functionals, confirm that our algorithm proficiently achieves the set goals, indicating that the amplitudes of the generated emission accurately reproduce the  angular correlations  and share the statistical properties of the employed CO targets.   {We thus aim at improving the current models of CO emission specifically in the high-Galactic latitude areas that have been hardly  observed by the most recent surveys, and, in doing so, to address and overcome the limitations affecting current models regions.} This research lays the groundwork for creating transformative synthetic simulations, leveraging convolutional neural networks tied to data procured from latest  observations.
\end{abstract}




\end{frontmatter}




\section{Introduction}
\label{introduction}

The  emissions of  Galactic foregrounds significantly contaminate the Cosmic Microwave Background (CMB) measurements. To address this challenge, recent efforts in the scientific community have focused on acquiring multi- wavelength observations, spanning   0.1 to 1 mm wavelength.  This multi-frequency approach enables accurate characterization of the spatial and spectral distributions of each foreground component, crucial for isolating the  cosmological signal. 

The sub-mm emission of our own Galaxy includes: thermal dust, synchrotron and molecular line emissions \citep{planck_2020}. Dust grains heated by optical starlight  emit thermal radiation in the Far Infra-red and microwave regime  ($\nu > 90$ GHz);   Galactic synchrotron radiation, due to cosmic electrons spiraling into the  magnetic field, mainly contributes at low frequencies ($\nu <70$ GHz). Both synchrotron and dust show  large levels of polarization ( up to 18\%, \citet{planck_2020}). The Carbon Monoxide (CO) emission in the sub-mm arises from the  molecular roational transitions, mainly the  $J :1-0,\, 2-1,\, 3-2$ lines.  One of the best CO survey to date is provided by \citet{2001ApJ...547..792D}.   { The survey probed the $J:1-0$ CO line along the Galactic mid-plane with resolution ranging from  7.5’ to 30’.} However, the    \citet{2001ApJ...547..792D} survey  was not planned to observe intermediate Galactic latitude regions as it requires longer integration times. On the other hand,      \emph{Planck}  released CO emission maps with similar level of accuracy but observations  far from the Galactic plane  ($|b|>30 \deg $ )  have been shown to be noise  dominated \citep{planck2013-p03a}. 
 
 CO molecular lines are expected to be polarized at lower fractional levels <1\% \citep{1981ApJ...243L..75G}.  \citet{Puglisi:2017},  showed how neglecting the polarized emission, mainly from the brightest rotational line of   $J:1-0$     may potentially  result into contaminating   the polarization measurements of future CMB experiments (e.g.\textit{ Simons Observatory (SO}, \citet{2019JCAP...02..056A,2022ApJ...929..166H}) that will further shed light into physics of primordial universe. To date, CO polarized emission has been observed in the brightest regions of molecular cloud complexes and around the Galactic plane, however, \textit{SO} observations will be instrumental   to set more stringent constraints over regions far from the Galactic plane.
 Because of the intrinsically dim fractional level of  CO polarization and the long integration time required to achieve a significant detection, it is difficult to carry out CO polarization surveys on wide areas in the sky   \citep{2001ApJ...547..792D}.  In the absence of data-based templates, the approach to modeling CO has been to assume a small degree of polarization applied to total intensity maps. 
 
 Moreover, to cope with the lack of data at high-Galactic latitudes,   \citet{Puglisi:2017} proposed a model of Galactic molecular clouds to   simulate the emission of CO lines in molecular clouds  and extrapolate it at high Galactic latitudes. Their model  is driven by   observational constraints related to the 3D  spatial distribution and to the intrinsic properties of the clouds, and it has been already employed in several contexts of Galactic science in the microwaves \citep{Hensley:2022,Borrill_2025}.

  {However, this approach has the drawback that it yields simulations which, although realistic in themselves, are completely uncorrelated with the other tracers commonly employed to monitor molecular clouds (e.g. HI and HII regions, free–free emission, thermal dust, etc.).}

Therefore, this work aims  at overcoming  the limitations presented in \citet{Puglisi:2017} by  leveraging the latest  developments of Generative algorithms in natural image processing.
 In Section~\ref{sec:data}, we present the data-sets and methodology to select the regions of interest, in Section~\ref{sec:arch} we describe the network architecture employed. Results are discussed in Section~\ref{sec:results} and in Section~\ref{sec:conclus} we derive the conclusions and future outlooks.

\begin{figure*}[!h]
	\centering 
	\includegraphics[width=1\textwidth]{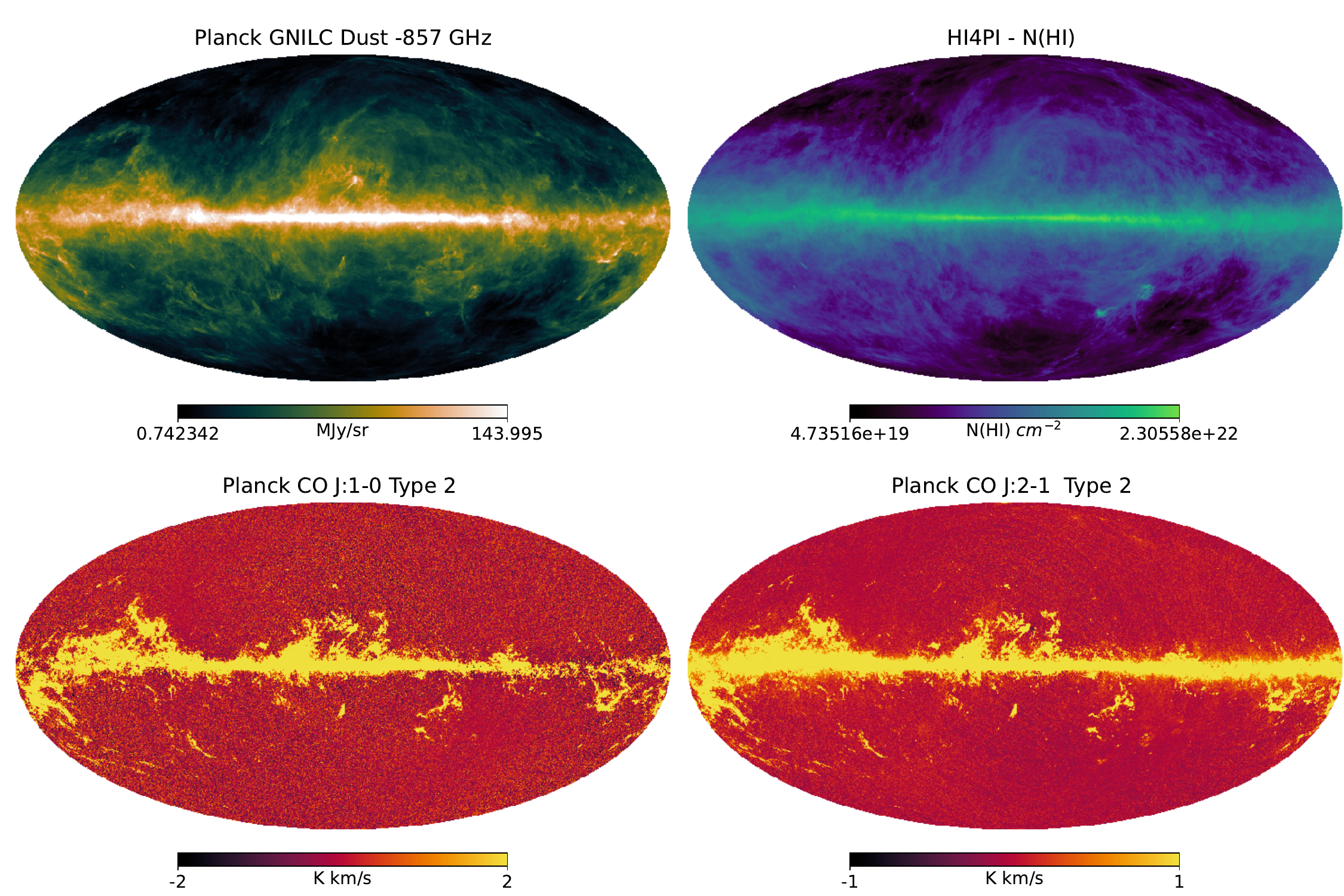}	
	\caption{(top panel) Fullsky maps  \emph{Planck} thermal dust at 857 GHz, $N(HI)$ column density map from HI4PI survey and  (bottom panel)\emph{Planck} CO \texttt{Type 2} $J:1-0$ and $J:2-1$ maps are shown respectively from the left to the right column.  } 
	\label{fig:inputs}%
\end{figure*}

\section{Data}\label{sec:data}

The selection of the training data set is primarily determined by exploiting the observed correlations between molecular emissions, the cold neutral atomic medium and the thermal emission of dust. Although recent literature \citep{Heyer2015} indicates that morphological correlations can effectively trace one emission as a function of the other, the relationships between them are highly non-linear and, to date, hard to express with an analytical relation.

The  \emph{Planck} satellite has released thermal dust intensity maps with an unprecedented sensitivity in the range of frequencies from 217-857 GHz.  We employ the Planck Generalized Needlet Independent Linear Component (GNILC)  2015 component separated map\footnote{\url{COM\_CompMap\_Dust-GNILC-F857\_2048\_R2.00.fits}} at $857$\,GHz \citep{planck2016-XLVIII}
 from the Data-release 3. The map was  obtained at the angular resolution of  $5^{\prime}$ with the GNILC method  aimed at optimally  separating the thermal Galactic emission from the extra-galactic contribution due to the CMB and the Cosmic Infrared Background (CIB). We employ the GNILC map at 857 GHz as it is the one with the highest SNR among the  \emph{Planck}  maps. 

 As molecular emission arises in the densest filaments and clouds emitting  HI atomic emission, we employed the $\rm N({HI})$ map from HI4PI survey \citep{hi2016} obtained by integrating data from the coverage of the \textit{Effelsberg-Bonn HI} Survey and the \textit{Galactic All-Sky Survey (GASS)}. The angular resolution   ($16.2^{\prime}$) and the improved sensitivity (43 mK) makes  it an unprecedented dataset with  complete spatial sampling of Galactic HI. Data have been projected onto a HEALPix \citep{Gorski:2005} map and made publicly available\footnote{\url{https://doi.org/10.26093/cds/vizier.35850041} }.
 
Galactic CO line emission  has been detected even in the broad photometric \emph{Planck} bands \citep{planck2013-p03a}. They  released the  \texttt{Type-2} templates\footnote{\texttt{HFI\_CompMap\_CO-Type2\_2048\_R2.00.fits}} based on a multi-frequency analysis in order to separate the first two rotational CO lines, $J:1-0$ and $J:2-1$, from the diffuse Galactic emission and the CMB. Although this approach is more prone to modeling errors in low SNR regions, we use these maps since they are employed in high SNR areas. The \texttt{Type-2} CO templates have a resolution of $15 ^{\prime}$, see Fig. \ref{fig:inputs}.

\section{\texttt{Cycle-GAN} model}\label{sec:arch}

\begin{figure*}
\centering 
    \includegraphics[width=1.2\columnwidth, angle =270, trim= 3.cm 0  3cm 0 , clip=true ]{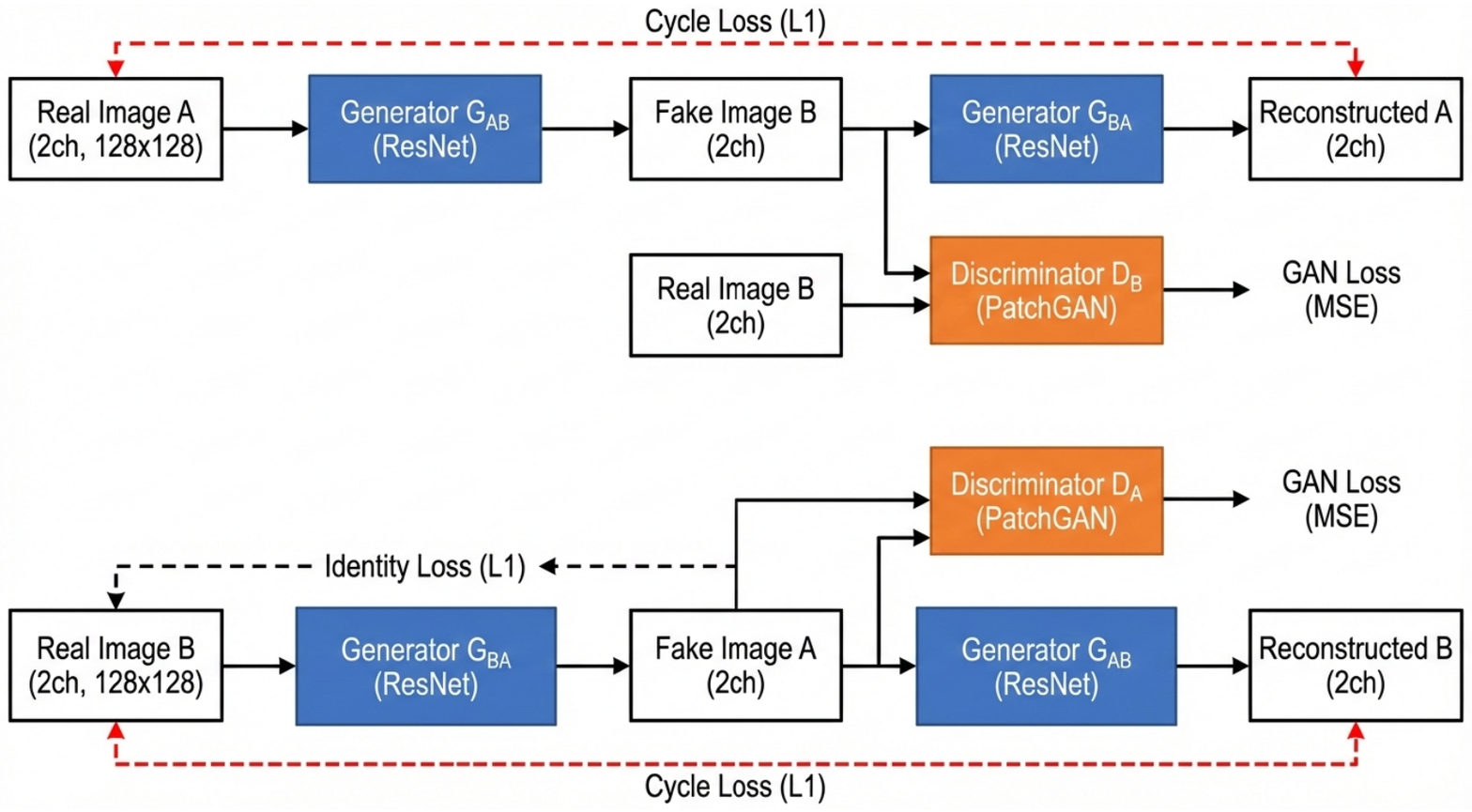}
    \caption{Overall architecture and workflow  of the \texttt{Cycle-GAN} implemented in this work. }\label{fig:arch}
\end{figure*}
  { \citealt{Goodfellow2014} is a widely used machine learning framework that has proven particularly effective in a variety of applications, most notably in image reconstruction tasks. In its standard formulation, a GAN can be understood as an \emph{adversarial interplay} between two neural networks: a \emph{generator}, $G$, and a \emph{discriminator}, $D$. The role of $G$ is to synthesize artificial samples that closely mimic the distribution of the training data, while $D$ is trained to assess each input and decide whether it originates from $G$ or is an authentic image drawn from the training set.}

{The objective function of the GAN is constructed so that $G$ is penalized when it produces images that appear unrealistic or that statistically diverge from the true data distribution, whereas $D$ is penalized when it incorrectly classifies a generated image as real or a real image as fake. Through this adversarial training process, the two networks are iteratively optimized until the generator becomes capable of producing high-quality images that the discriminator can no longer reliably distinguish from genuine samples. } 

  {Lately, GANs have been applied in a wide range of scenarios and for various purposes, consistently demonstrating their ability to generate realistic astrophysical datasets (see \citet{Puglisi_2020,2021ApJ...911...42K, Yao:2024,ullmo2025}). Motivated by these successful results, we choose to base our approach in this work on such methodologies.}

\citet{zho2020}  developed a method that can learn to translate images from one style or a ``collection'' to another without needing paired examples. This is a significant breakthrough in unpaired image-to-image translation, a field that aims to convert an image from one representation to another.
Traditional image translation methods rely on supervised learning, where a system is trained on thousands of matching input-output pairs (e.g. a photo and its corresponding grayscale version). On the other hand, unsupervised methods like Generative Adversarial Neural networks (GANs) method does not  guarantee a meaningful relationship between the input and the translated output and often leads to \emph{mode collapse}, where the system would produce a single, generic output image for all inputs.
 
  {To overcome this, \citet{zho2020} added a key constraint: the \emph{cycle consistency}, by training two models simultaneously, each one accounting for a  generator and an adversarial discriminator.  Namely: 
\begin{itemize}
\item a generator $G_{AB}: A\rightarrow B$ translating images from domain $A$ to domain $B$ and the discriminator $D_B$ judging whether a translated image from $G_{AB}$ looks like a real image from domain $B$; 
\item  a generator  $G_{BA}: B\rightarrow A$ translates images from domain $B$ to domain $A$, coupled with the discriminator  $D_A$, ensuring the  images generated by $G_{BA}$  look like the ones  in domain $A$. 
\end{itemize}
}

Details on the  generator and discriminator architectures can be found in \ref{app:arch}.

  {In our implementation, we choose  as adversarial loss the \emph{Mean-Squared Error}  (MSE) loss   instead of the standard \emph{binary Cross-Entropy}, this is mainly due to the fact that the former has shown to be more stable during training \citep{isola2018,zho2020}.  So that  for $D_B$ and for $G$ the loss functions are  defined as follows:}
 \begin{align}
     \mathcal{L}_{GAN} (  D_B  )=&\,  \frac{1}{2}  \mathbb{E}_{b\sim p_{data}(B)}  \left[ \parallel  D_B(b) -1 \parallel ^2  \right] \\
     +&\frac{1}{2}\mathbb{E}_{a\sim p_{data}(A)}  \left[ \parallel  D_B(G_{AB}(a)  \parallel ^2  \right] , \\
     \mathcal{L}_{GAN} (G_{AB})=&\,    \mathbb{E}_{a\sim p_{data}(A)}  \left[ \parallel  D_B(G_{AB}(a)) -1 \parallel ^2  \right] 
     \label{eq:adv_loss}
 \end{align}
   {with $p_{data} $ being the data distribution of the training samples.  $G_{AB}$ aims to minimize this loss against the adversarial discriminator $D_B$ that instead tries to maximize it.  }
 
   {From eq.\eqref{eq:adv_loss}, we can observe that the adversarial training process reaches convergence when the discriminator can no longer  differentiate between real images and those produced by the generator. In practical terms, this situation corresponds to the point at which the adversarial loss stabilizes around a value of approximately $\sim 0.25$, indicating that the discriminator’s predictions have effectively become indistinguishable for real and generated samples.  The adversarial loss for $G_{BA}$ and $D_A$ are similarly  defined.}

  Moreover, the \texttt{Cycle-GAN} objective function    is  composed  also by other two terms : 
\begin{itemize}
    \item a cycle-consistency loss  $\mathcal{L}_{\rm cyc}$: 
 
 \begin{align*}
     \mathcal{L}_{\rm cyc}(G_{AB},G_{BA}) =&\,  \mathbb{E}_{a\sim p_{data}(A)}\left[ \,\parallel G_{BA}(G_{AB}(a) )  - a \parallel_1 \right] \\
     +& \,\mathbb{E}_{b\sim p_{data}(B)}\left[ \, \parallel G_{AB}(G_{BA}(b))   - b \parallel_1  \right], 
 \end{align*}
ensuring the \emph{forward cycle consistency}, i.e. for each image $a$  when we apply consecutively $G_{AB}$ and $G_{BA}$ we retrieve  back  the original image,  i.e.
\begin{displaymath}
    a \rightarrow G_{AB}(a) \rightarrow G_{BA}(G_{AB}(a)) \approx a. 
\end{displaymath}

Similarly, 
for each image $b$ from domain $B$ the  \emph{backward cycle consistency} grants the reverse process.  We adopt  $L_1$ norm for $\mathcal{L}_{\rm cyc}$. 
\item an identity loss, $\mathcal{L}_{\rm id}$:
\begin{align*}
     \mathcal{L}_{\rm id}(G_{AB},G_{BA}) =&\,  \mathbb{E}_{a\sim p_{data}(A)}\left[ \,\parallel G_{BA}(a )  - a \parallel_1 \right] \\
     +& \,\mathbb{E}_{b\sim p_{data}(B)}\left[ \, \parallel G_{AB}(b)   - b \parallel_1  \right], 
 \end{align*}
 ensuring that if a generator is given an image already belonging to the target domain, it should leave it unchanged. 
\end{itemize}
 The relative importance of the two objectives is controlled by two hyper-parameters, as it follows: 
\begin{align*}
     \mathcal{L}_{\rm total} &=  \mathcal{L}_{GAN}(G_{AB} )+ \mathcal{L}_{GAN}(G_{BA} )\\
            &+ \lambda_{\rm cyc} \mathcal{L}_{\rm cyc}(G_{AB},G_{BA}) \\
            &+ \lambda _{\rm id} \mathcal{L}_{\rm id}(G_{BA},G_{BA}),   
 \end{align*}
with $\lambda_{\rm cyc}=10$ and $\lambda_{\rm id}=5$ (values chosen following the recommendations from  \citet{zho2020}).  Further  details on the architecture of the generator and discriminatror can be found \ref{app:arch}.
  
\subsection{Training}\label{sec:prep}

To construct the dataset and facilitate the training process, we convolve the four input maps to the coarsest angular resolution of $16 ^{\prime} $ and make sure that all the maps are pixellated following the HEALPix scheme, with the same pixel  resolution of $\sim 1.7^{\prime}$, corresponding to an \texttt{nside}=2048 HEALPix grid. 

  {One of the advantages of GANs is to employ unpaired data sets, i.e. training with an  intentionally mismatched data from $A$ and $B$ domains, in order to let the network learns the domain translation rules, not just how to copy one specific image to another.  This is particularly remarkable when dealing with realistic datasets, where observations could be highly affected by noise or could involve different kinds of emissions.}

As can be seen in Fig.\ref{fig:inputs}, both the \emph{Planck} 857 GHz dust and the \emph{HI4PI} maps are signal-dominated, with a signal-to-noise ratio (SNR) > 3 across the full-sky coverage \citep{planck_2020,hi2016}. This high SNR ensures that the features we use as inputs are reliably measured and noise, albeit present, is subdominant with respect to the signal.

To construct the input data, we therefore segmented the \emph{Planck} and \emph{HI4PI} all-sky maps into tiles of size $3\times 3 \deg^2$, each tile consisting of $128\times 128$ pixels, accounting for a total of 8,452  patches.

\begin{figure}[h!]
	\centering 
	\includegraphics[width=.5\textwidth]{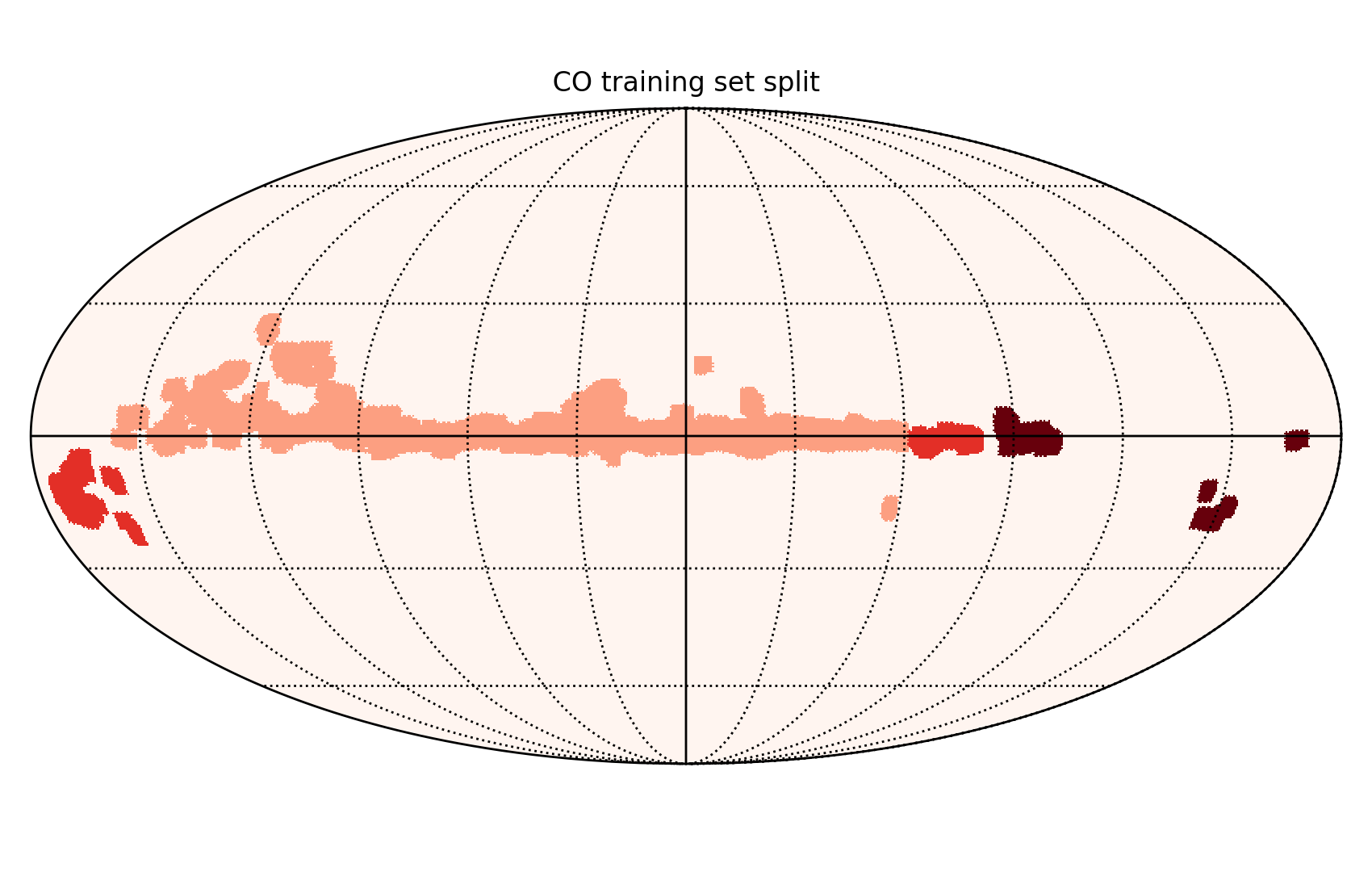}	
	\caption{High SNR regions of CO emission from the \emph{Planck} \texttt{Type-2} maps. Different colors refers to patches employed for (orange) training, (red) validation, (brown) test sets. } 
	\label{fig:split}%
\end{figure}

On the opposite,  the \emph{Planck} CO \texttt{Type-2} maps  (the output targets)   are  more   contaminated by noise especially far from the Galactic mid-plane. We, thus,  consider $3\times 3 \deg^2 $ tiles, made of $128\times 128$ pixels, chosen from SNR> 8 regions  as  illustrated in Fig. \ref{fig:split}. 

  {The dataset is  further augmented through  flipping, aiming to further enhance the size and diversity of the training dataset,  leading  to a  total of 5,205 (whose  336 left for testing) and 22,818  respectively for the output and the input maps. We left  336  paired tiles for  testing  to allow a more direct comparison of the performances.  }

We remark here that  the SNR >8  selection of  regions may correspond to   systematically select optically thick regions, notably near the plane.
We, thus,make sure to construct the whole training data set  including also   optically thin regions at intermediate galactic latitude or in the outer galaxy, allowing the network to inspect those emission regimes. 

  {In fig.\ref{fig:split}, we show in different colors  the regions selected for training, validation and test. We notice that each data set employs tiles selected from the Galactic plane and from the outer Galaxy or intermediate Galactic latitudes.  }Additionally, we mitigated any overlap between tiles to preserve morphological diversity within each  set. This leads to a total of   1,733  tiles (whose locations are shown in Fig.~\ref{fig:split}).

  {To compress the dynamic range, we firstly   \emph{log-}transform each HEALPix map, $M$, i.e.   :
\begin{displaymath}
    m =  \log_{10} ( M+\epsilon ), 
\end{displaymath}
with $\epsilon =10^{-6}$, appropriately chosen to avoid \emph{NaN errors}   with zero values in $M$. We also make sure to shift to 0 any negative values   due to instrumental noise. We, then,   \emph{minmax} normalized the map so that it globally ranges   in the interval $[-1,1]$ : }
\[ \tilde{m} = 2 \frac{(m - \min(m))}{\max(m) - \min(m) } -1 . \]

We, finally,  proceed to build the training set by projecting into squared patches from the normalized map $\tilde{m}$. 

  {We consider  batches of 32 tiles and perform the training onto  an  NVIDIA A100-SXM4-40GB  GPU node  of the  supercomputing facility PERLMUTTER at NERSC. The optimizer adopted is the \emph{Adaptive Moment Estimation} (ADAM) optimizer with a learning rate of $0.0002$ and $\beta$ parameters set to $(0.5, 0.999)$. }

\begin{figure}
    \includegraphics[width=1\columnwidth]{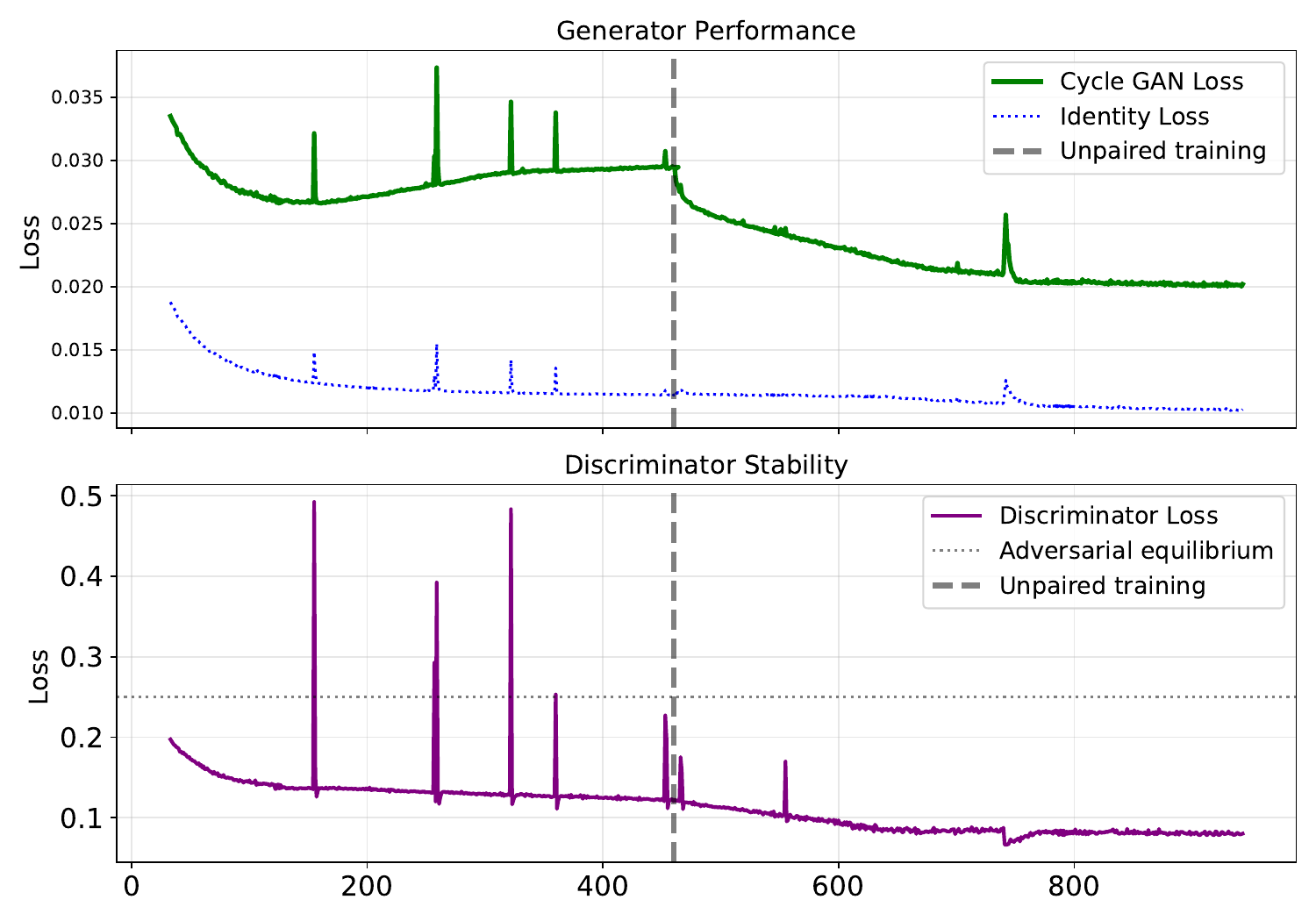}
    \caption{(top) Generator cycle (solid green) and identity (dotted blue) losses as a function of epochs. (bottom) Discriminator loss as a function of epochs. The vertical dashed thick line at epoch 460 marks the point  where the training switched from   paired   to unpaired. }\label{fig:losses}
\end{figure}
  {The training  employes 940  epochs and take $\sim 47$ hours (each one taking on average  3 min, with  a 0.1\%  spent for data loading, and 14\% for  batch processing time).}

  {We decide to perform the training with  a paired data set  to allow  the network learning  astrophysical correlations present in both input and output  features.    
As shown   in Fig.\ref{fig:losses}, , we notice that  after few hundred epochs, the generator loss starts to increase and reaches   a plateau of 0.03 after 460 epochs. This is expected since  the paired data set is by construction smaller  than the unpaired one (due to the limited SNR>8 CO regions). We, thus,  decide to switch into unpaired training by further   learning the features in both optically thin and optically thick regimes. During unpaired training, we increased also $\lambda_{\rm cyc}=20 $, to give more weight to the cycle loss.  
This  results in a sensible decrease of  the cycle  loss until it reaches the desired threshold of convergence at   $0.02$.   }

The code  has been made publicly available\footnote{\url{https://github.com/giuspugl/COnet.git} } together with the training weights\footnote{
\href{https://drive.google.com/drive/folders/12Ryc7K10PXpXlK2pCXWio-6rGCqB_lT9?usp=drive_link}{\texttt{Cycle-GAN} weights } } for the sake of reproducibility.

\section{Results}\label{sec:results}

\begin{figure*}[!h]
	\centering 
	\includegraphics[width=1.\textwidth, clip=true,trim=0 0cm 0 4cm]{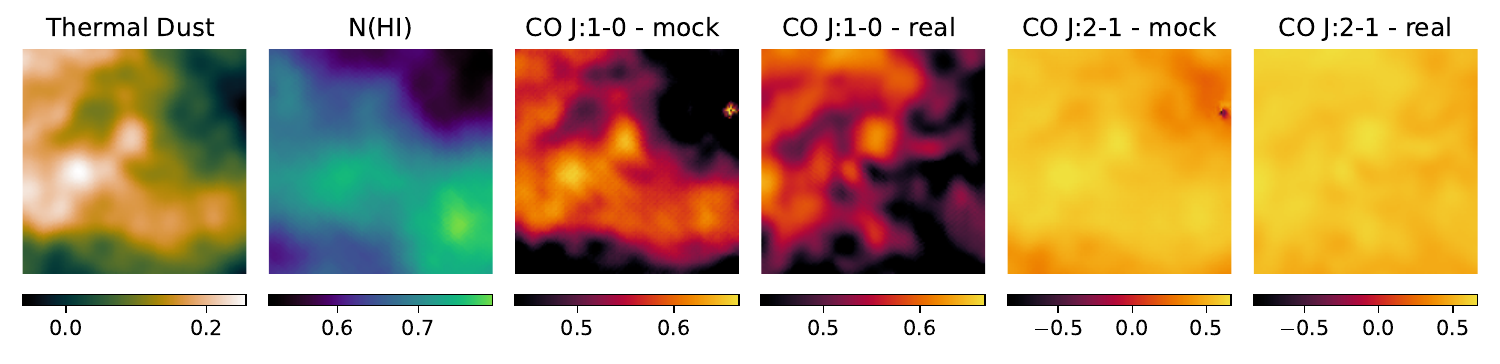}	
	\includegraphics[width=1.\textwidth, clip=true,trim=0 0cm 0 5cm]{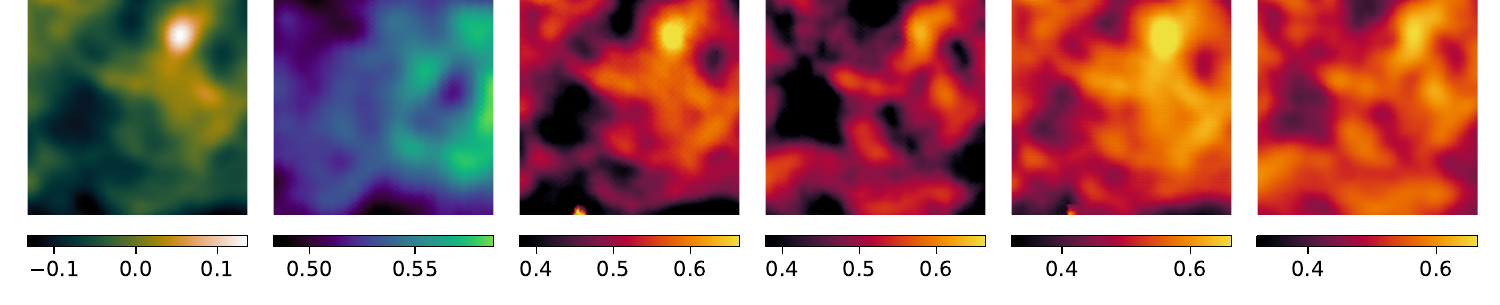}	
	\includegraphics[width=1.\textwidth, clip=true,trim=0 0cm 0 5cm]{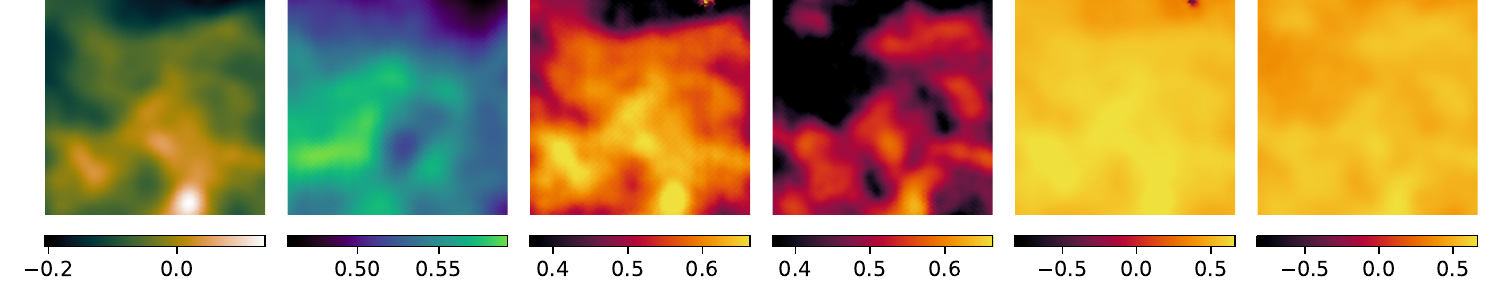}	
	\includegraphics[width=1.\textwidth, clip=true,trim=0 0cm 0 5cm]{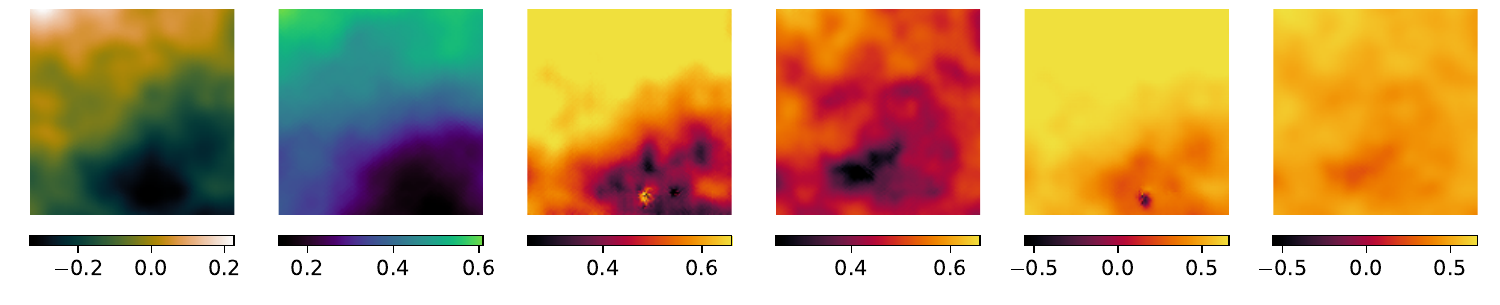}	
	\includegraphics[width=1.\textwidth, clip=true,trim=0 0cm 0 5cm]{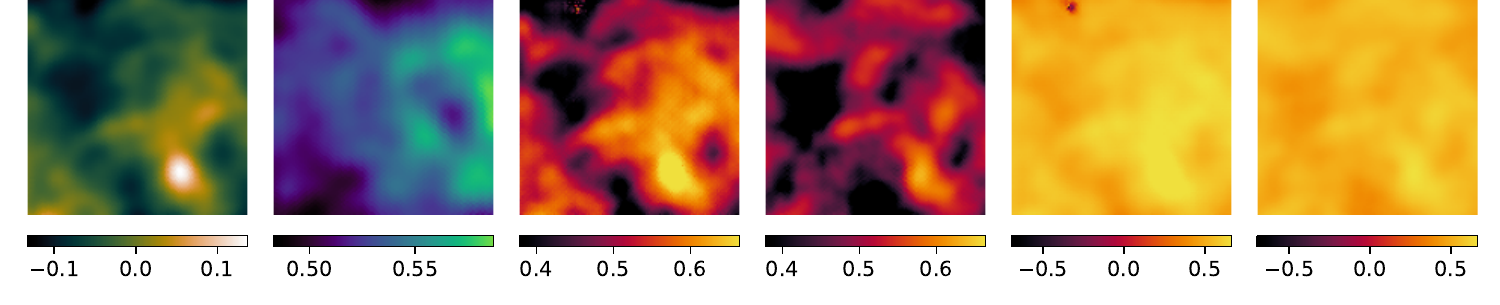}	
	\caption{$3\times 3 \deg^2$ tiles employed in the test set for training  the \texttt{Cycle-GAN} selected from  regions shown in Fig. ~\ref{fig:split}. \emph{Planck} thermal dust at 857 GHz,  $ N(\rm{HI}) $ column density map from HI4PI survey, \emph{Planck} CO \texttt{Type 2} $J:1-0$ and $J:2-1$ maps are shown respectively from the left to the right column, together with the predictions obtained with the \texttt{Cycle-GAN}. } 
	\label{fig:outputs}%
\end{figure*}
In  Fig.~\ref{fig:outputs}, we show a visual comparison of the patches generated with the \texttt{Cycle-GAN} against the ground truth in the test set. We notice that the network is able to reproduce similar filamentary and point-source structures as solely inspecting the input features represented by the dust and HI maps from the test set.

  { We can clearly state that the network consistently predicts  not only realistic  morphological shapes  but also coherent range of  amplitudes.}

  {To better guide the reader, we set the colorbars of the mock images to be the same as the real ones.   Moreover,  we notice the morphologies are mainly guided by the    the thermal dust input features. This outcome is, to some extent, expected, as astrophysical observations indicate that molecular emission closely correlates with the emission from cold dust. \citep{cosmoglobe2026}}.

\begin{figure*}[!h]
	\centering 
	\includegraphics[width=.49\textwidth ]{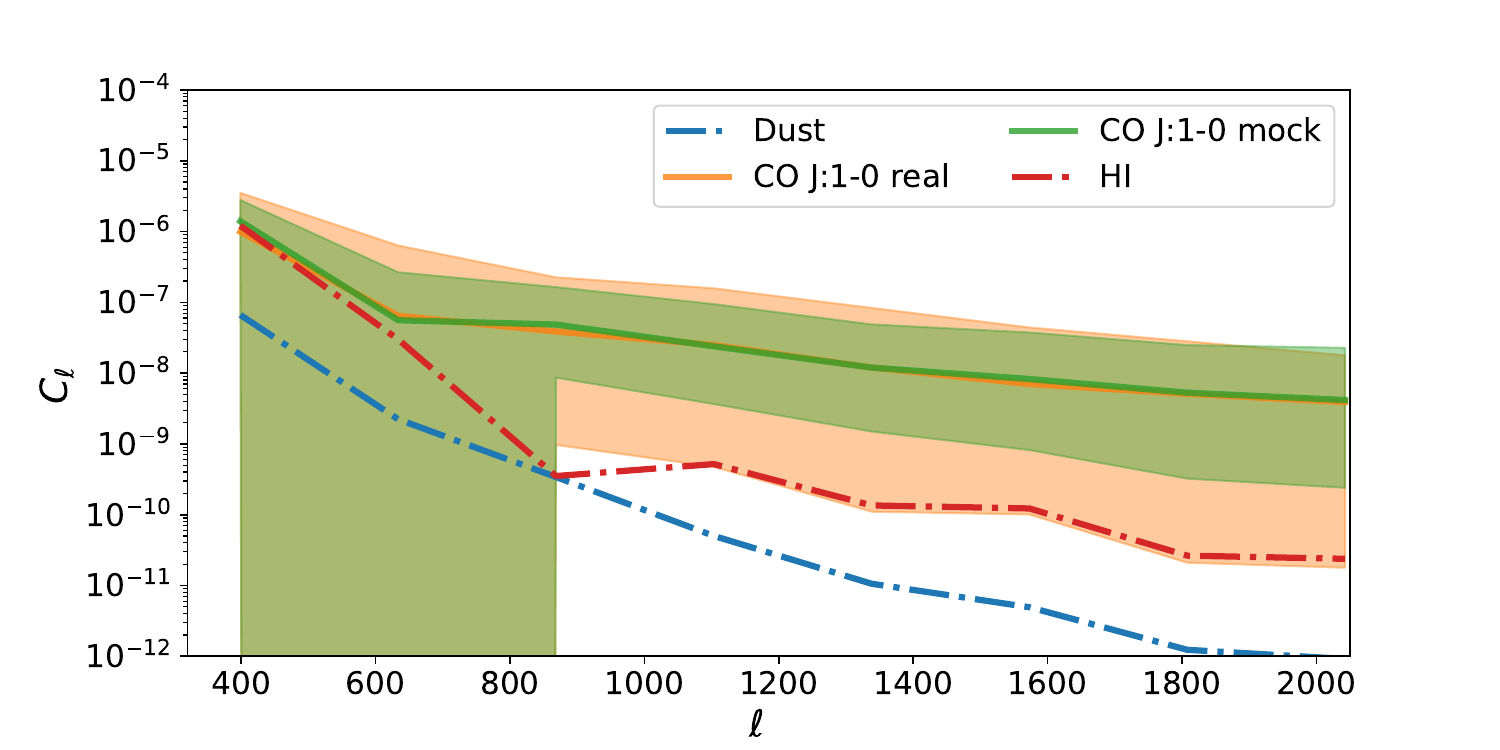}	
	\includegraphics[width=.49\textwidth ]{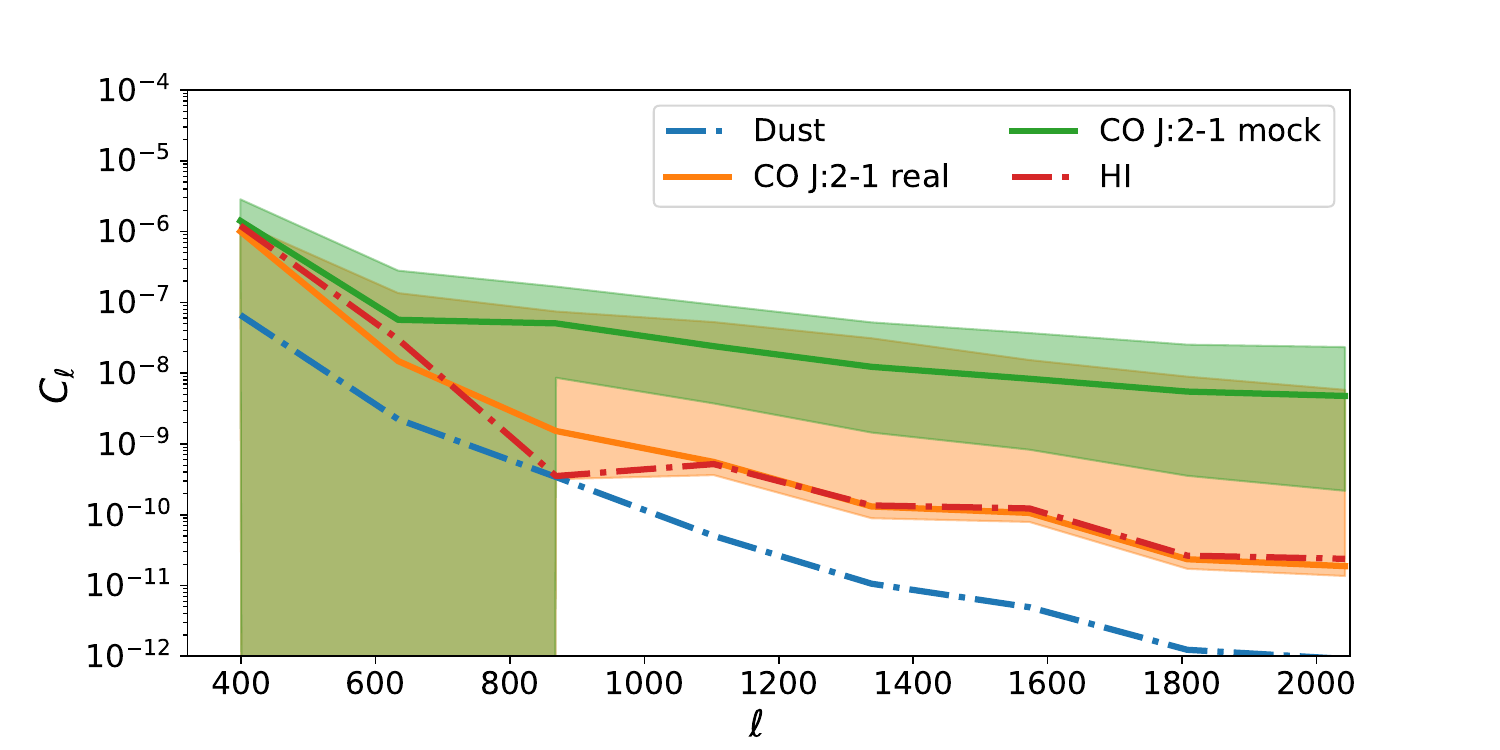}		
	\caption{ 
 Power spectra estimated from the test set.  Median  power spectra are shown as (solid orange) and (solid green)   respectively for the ones estimated from real and synthesized (mock)  CO $J:1-0$ (left panel) and $J:2-1$ (right panel) emissions.    The  green and orange shaded areas corresponds to the  $2\sigma$ interval of power  spectra estimated  respectively from  the whole  mock and real test set. Power spectra of input  dust and HI maps are respectively shown for comparison in  (dot-dashed blue) and (dot-dashed red). The spectra are uniformly binned with $\Delta \ell =200$. } 
	\label{fig:spectra}%
\end{figure*}

We then proceed with a more quantitative estimate by means of  the 2-point statistics, to assess  predicted data quality in terms of angular correlations in the synthesized images by  estimating the angular power spectrum of each tile of the test set   using \texttt{NaMaster} software \citep{Alonso:2019}. For most of Galactic emission,  we   expect the power spectrum  to be scale-invariant, following a power-law scaling in terms of the multipole $\ell$ as shown in Fig.\ref{fig:spectra}. 

As our patches involves  a small portion of the sky, where curvature effects are negligible, we can  run  \texttt{NaMaster} with  \textit{Flat-Field} estimators as recommended in the code documentation\footnote{\url{https://namaster.readthedocs.io/en/latest/api/pymaster.utils.html}}.    We further consider a  window function to  apodize each tile  with the \texttt{C2} profile, defined as follows: 
\begin{displaymath}
     w(x) =
     \begin{cases} 
     \frac{1}{2 } \left( 1 - \cos(\pi x) \right) , & \text{if } x < 1 \\ 1 & \mathrm{otherwise}
     \end{cases} 
\end{displaymath}
with $x = \sqrt{(1- \cos \theta ) /(1- \cos \theta_* )}$, $\theta$ being the distance between two pixel and $\theta_*= 0.5 \deg$. The power spectra are subsequently estimated with equally spaced bins of multipoles characterized by $\Delta \ell = 200 $.

To facilitate the comparison of spectra derived from maps with differing units of measure, we estimate the spectra  directly on the  \emph{log-normalized }  data set, as described in Subsection~\ref{sec:prep}.

 Fig.~\ref{fig:spectra} shows the power spectra estimated from the test set. We notice that   the   median (solid green ) from \texttt{Cycle-GAN} CO predictions almost follows the same scaling as   the ground-truth one (solid orange).  
 We remark here that  the estimated  power spectra  accurately reproduce the scaling across various angular scales, ranging from $0.5 \deg$ to $3 ^{\prime} $. 
 
 Moreover, we clearly are able to distinguish that the network is reproducing more the scaling  observed in the CO power spectrum  than  the dust and HI ones. We  also  show the $2\sigma $ excursion range of  the power spectra estimated from the different tiles of the predictions and ground-truth maps.   {As both the  excursion range overlaps in most of the angular multipoles, we can quantitatively assess that the angular correlations of the mock and real data are consistent withing the $2\sigma$ interval. }

\begin{figure*}[!h]
	\centering 
	\includegraphics[width=1.\textwidth,  ]{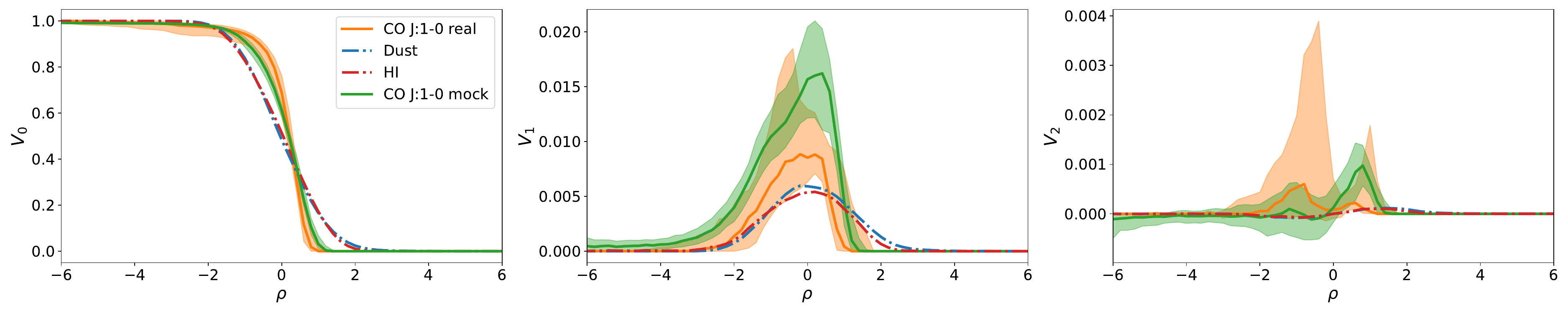}	
	\includegraphics[width=1.\textwidth,  ]{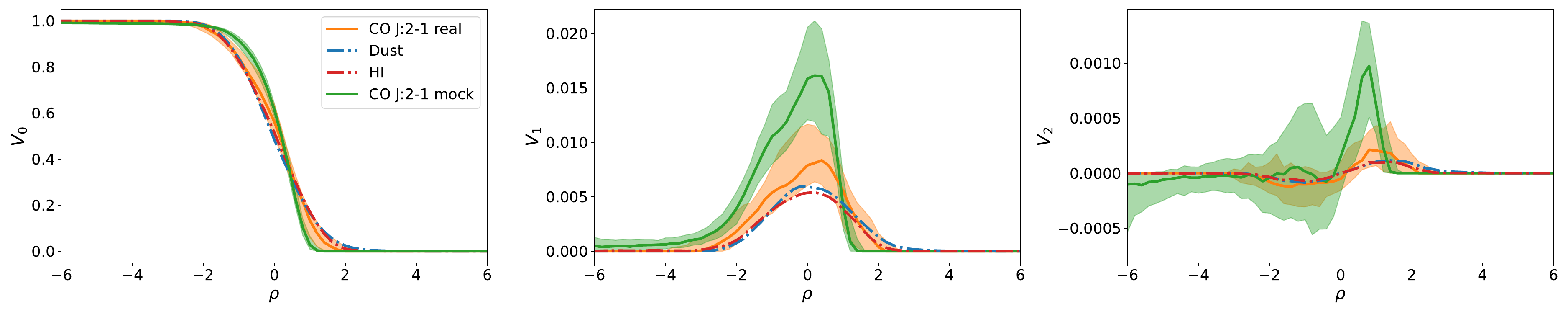}		
	\caption{From left to right, ${V_0, V_1, V_2}$ MFs estimated from the test set for the $J:1-0$ and $J:2-1$  CO median  emission from the ground-truth (solid orange) compared with the mock one syntethized by the \texttt{Cycle-GAN} (solid green). As in Fig.~\ref{fig:spectra}, we also show the  $2\sigma$  interval  of the excursion range as  shaded orange and  green respectively for the ground-truth and mock sets. For the sake of comparison we also show the median MFs of the input dust (dot-dashed blue) and the log of HI column density (dot-dashed red). } 
	\label{fig:mink}%
\end{figure*}

\begin{figure*}[!h]
	\centering 
	\includegraphics[width=1.\textwidth, clip=true,trim=0 0cm 0 0cm]{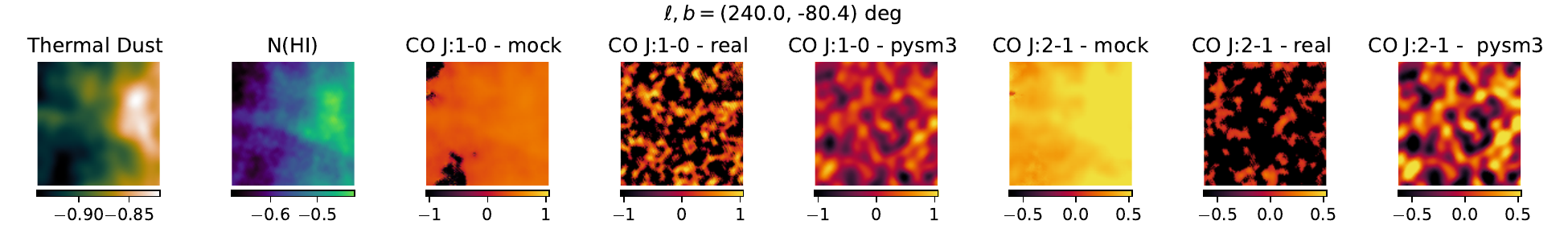}	
	\includegraphics[width=1.\textwidth, clip=true,trim=0 0cm 0 0cm]{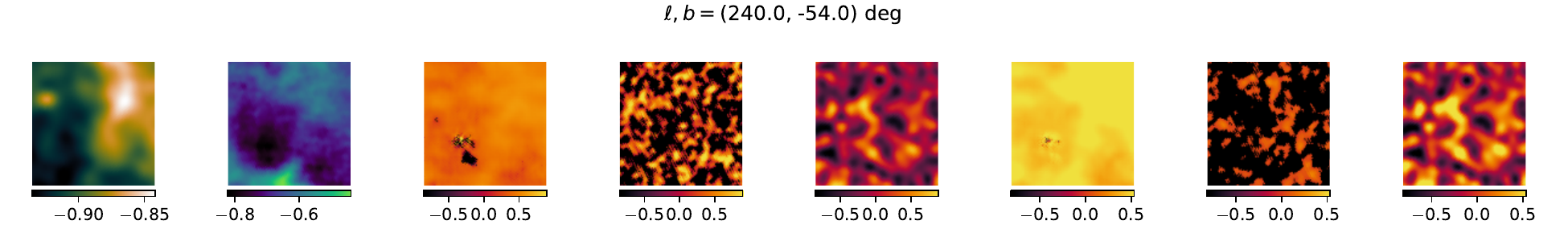}	
    \includegraphics[width=1.\textwidth, clip=true,trim=0 0cm 0 0cm]{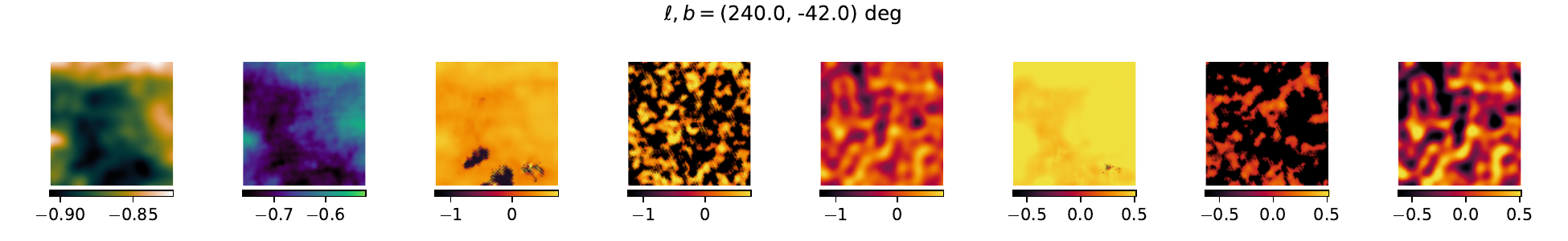}	
      \includegraphics[width=1.\textwidth, clip=true,trim=0 0cm 0 0cm]{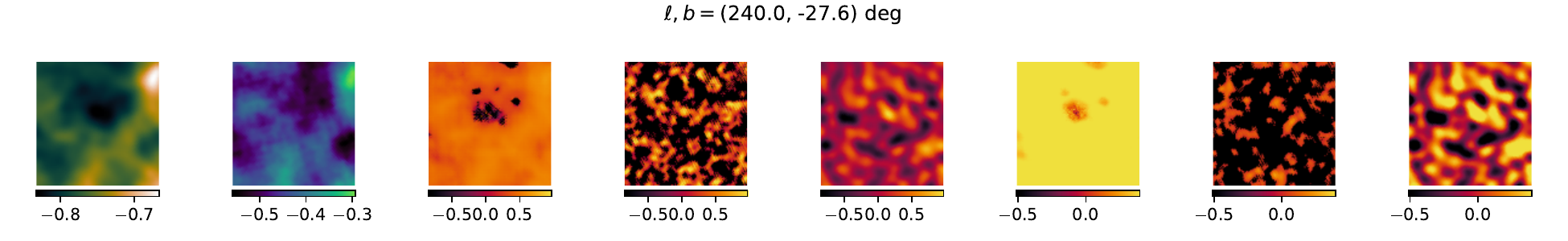}	
            \includegraphics[width=1.\textwidth, clip=true,trim=0 0cm 0 0cm]{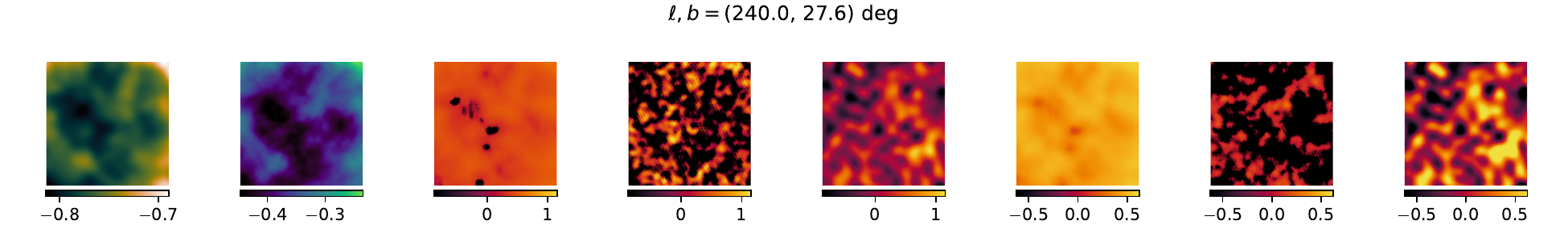}	
    \includegraphics[width=1.\textwidth, clip=true,trim=0 0cm 0 0cm]{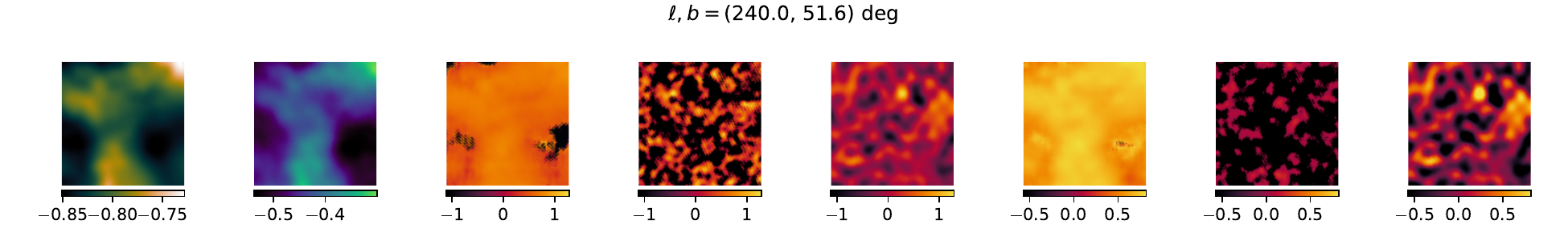}	
    \includegraphics[width=1.\textwidth, clip=true,trim=0 0cm 0 0cm]{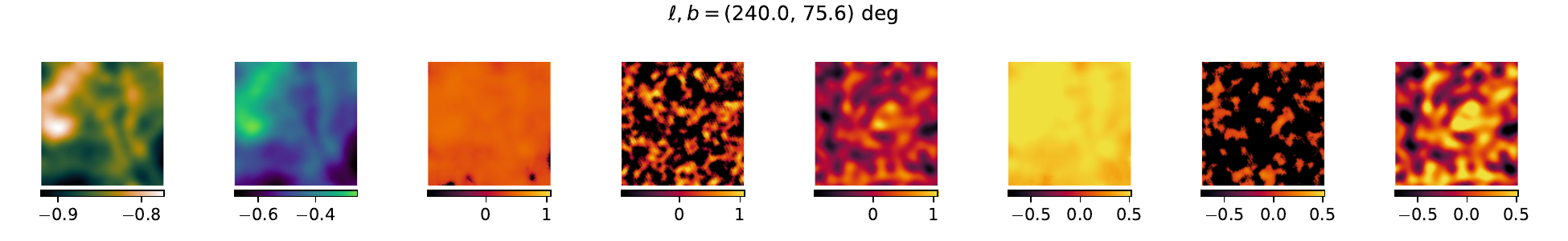}	

	\caption{{$3\times 3 \deg^2$ tiles employed for application of  the \texttt{Cycle-GAN} at high-Galactic latitudes. From left to right columns we show maps (in log-normalized units) : \emph{Planck} thermal dust at 857 GHz,  $ N(\rm{HI}) $ column density map from HI4PI survey,  CO  $J:1-0$  and  $J:2-1$ from mock, \emph{Planck} \texttt{Type 2},and \texttt{pysm3} model maps. To facilitate the comparison  between the CO emission maps, we set their colorbars to the same range.  }  }
	\label{fig:applic}%
\end{figure*}

We further  assess through the Minkowski Functionals \citep[MFs,][]{Minkowski1903}   the  degree of non-Gaussianity  in the mock CO emission compared with the ground-truth one.  MF  have been widely employed in the literature to estimate higher-order statistical properties of maps \citep[e.g.,][]{Puglisi_2020,Krachmalnicoff:2021, Martire:2023,duque2023,Carones:2024,Yao:2024, Borrill_2025}. 

For a given  $n$-dim set  whose range is defined with a threshold $\rho$,  there are $n+1$ MFs that could provide high-order statistics informations \citep{hadwigerVorlesungenUeberInhalt1957}. In this work we employ $n=2$ maps, thus we utilize 3 MFs: ${V}_0$, ${V}_1$, and ${V}_2$, representing  respectively the area, perimeter, and connectivity as a function of the threshold $\rho$. 
We estimate MFs with  the package  \texttt{pynkowski}\footnote{\url{https://github.com/javicarron/pynkowski}} \citep{duque2023}. 

Fig.~\ref{fig:mink} shows the MFs  estimated from the  patches in the test set, once they are standardized (zero mean and  variance equal to 1). We employ the same color scheme as in Fig~\ref{fig:spectra}:  the mock  (solid green) and ground-truth  (solid  orange) emissions,  accounting for the  $2\sigma$ interval of the excursion range respectively shown  in shaded green and orange. 
We remark that the MFs are consistent within the  $2\sigma$  interval, indicating that also the high-order statistics has been properly learnt and reproduced  by the \texttt{Cycle-GAN}.  

By looking at  the MFs estimated  dust and HI patches,  we clearly see a remarkable difference when compared with the  ones estimated from  mock data. This further points  to the fact that the high-order statistics of generated mock data reproduces the one observed in real ones.

\subsection{Application at high-Galactic latitudes}

In this Sub-section, we assess the performances of our trained \texttt{Cycle-GAN} to generalize in regions at high-Galactic latitudes, where current CO data are currently lacking or are strongly affected by noise. 

We therefore divide the full-sky dust and HI observations into 8,452 $3\times 3 \, \mathrm{deg^2}$ tiles performing the same  preprocessing steps followed in Sub-section\ref{sec:prep} and run the \texttt{Cycle-GAN} to predict on   this data set. 

In Fig.\ref{fig:applic},  we show a subset of these predictions, selected from all the patches at the same Galactic longitude $\ell=240^{\circ}$ and at several Galactic latitudes. This choice is specifically driven by the fact that the training set does not encode data along this longitude, as can be noticed from Fig.\ref{fig:split}. 

We compare the predictions against the \emph{Planck} CO \texttt{Type-2} and we clearly state that for all the patches shown in Fig.\ref{fig:applic} the latter are strongly contaminated by instrumental noise in both the two CO  lines, 

As stated above,  we aim to compare our predictions against the Monte-Carlo Molecular 3D (\texttt{MCMole3D}) model from \citet{Puglisi:2017}. Their model populated high Galactic latitude CO clouds  by fitting several parameters from  low latitudes   CO emission data. A notable distinction from this approach is that they use   \emph{Planck} CO \texttt{Type-1} maps as templates for the  emission  of  both $J:1-0$ and $J:2-1$ lines. 
Recently, the MCMole3D maps have been incorporated into the Python Sky Model 3 (\texttt{pysm3}, \citet{Borrill_2025})\footnote{See the   model  labeled as \texttt{co3}  in the \href{https://pysm3.readthedocs.io/en/latest/models.html\#co-line-emission}{pysm3 documentation}.}.
We, therefore, run the \texttt{pysm3} to simulate the  CO $J:1-0$ and $J:2-1$ emission from the  \texttt{co3}  model, log-normalized and projected into square tiles.  

 When comparing  our mock maps  with the \texttt{pysm3} ones, we clearly notice several differences among the two data sets. 
 
 On one hand, being  the \texttt{pysm3} model   based on the \emph{Planck} CO \texttt{Type-1} maps, its emission  is largely contaminated  by the instrumental noise. This is expected, as the CO emission is expected to be weaker at high-Galactic latitudes, making instrumental sensitivity a critical factor. Indeed, the \texttt{Cycle-GAN} predicts pixel values of -1 in the areas, that are likely to be associated with intrinsically faint emission and more prone to be contaminated by instrumental  noise.
 On the other hand, the morphology of the emission in our predicted maps more closely reproduces the diffuse, filamentary structures observed in dust and HI, whereas in the \texttt{pysm3} maps, residual noise   readily distorts the morphology, favoring compact, point-like emission features at the expense of the extended structures.

\section{Conclusions}\label{sec:conclus}
In this work, we  introduce a novel  methodology  utilizing a \texttt{Cycle-GAN} implementation  to predict  realistic non-Gaussian sub-degree CO emissions. 
Our algorithm is employed to learn molecular CO emissions by examining inputs from thermal dust emissions and atomic emissions while targeting CO emissions in the two most prominent rotational transitions, $J:1-0$ and $J:2-1$. The network is trained to generate CO emissions specifically in regions where observations  have been achieved with  SNR>8. The data set encodes square sky regions with a physical size of $3\times 3 \deg^2$ and a resolution of $128\times 128$ pixels. 

  {The  quality of the results, quantified through the power spectra and MFs, demonstrate that our algorithm successfully meets the aim of this work, illustrating that the angular correlations of the generated features properly  scale with angular size and have the same high-order statistics as the CO targets employed. This has been evaluated by assessing consistency within $2\sigma$ with a comparison of power spectra and MFs of the mock and real data sets. }

  {To further corroborate the performance of our method, we show that that our methodology is able to generalize at high Galactic latitudes which have not been taken into account in the training set. This is particular of interest as molecular complexes far from the Galactic plane are characterized by low emissivity and low opacity, making  surveys difficult to carry out in those regions. We further compare the predictions with a  model from the \texttt{pysm3} package commonly adopted by the community.  From such a comparison, we expect  differences not only in the amplitudes but also in the morphologies, because the \texttt{pysm3} model is more affected by the \emph{Planck } instrumental noise, whereas the model presented in this work  completely relies on  high SNR of thermal dust and HI maps.   }  

This work paves the road for a new Galactic CO emission model at high Galactic latitudes that will be presented in a future work. 
More remarkably, it demonstrates the advantages of employing such algorithms to cope the  current limitations arising from the available observational data sets.

\section*{Acknowledgements}
This work has been supported by Italian Research Center on High Performance Computing Big Data and Quantum Computing (ICSC), project funded by European Union - NextGenerationEU - and National Recovery and Resilience Plan (NRRP) - Mission 4 Component 2 within the activities of Spoke 3 (Astrophysics and Cosmos Observations).
This research used resources of the National Energy Research Scientific Computing Center, a DOE Office of Science User Facility supported by the Office of Science of the U.S. Department of Energy under Contract No. DE-AC02-05CH11231 using NERSC award HEP-ERCAP0032657.
GP acknowledges financial support under the National Recovery and Resilience Plan (NRRP), Mission 4, Component 2, Investment 1.1, Call for tender No. 104 published on 2.2.2022 by the Italian Ministry of University and Research (MUR), funded by the European Union – NextGenerationEU– Project Title ``SHIFT'' – CUP 55723062008 - Grant Assignment Decree No. 962 adopted on June 30th 2023 by the Italian Ministry of University and Research (MUR).

\appendix

\section{\texttt{Cycle-GAN} Architecture } \label{app:arch}

The architecture shown in Fig.\ref{fig:arch} consists of two generator models and two discriminators. The cycle consistency loss is the core component that ensures successful transformation from one domain to another and back again, thus maintaining the original content. We remark here, that we employ the exact same architecture for both discriminators $D_X$ and $D_Y$ and for the generators, $G$ and $F$. 

\begin{figure*}
\centering
    \includegraphics[width=1.5\columnwidth, angle=270] {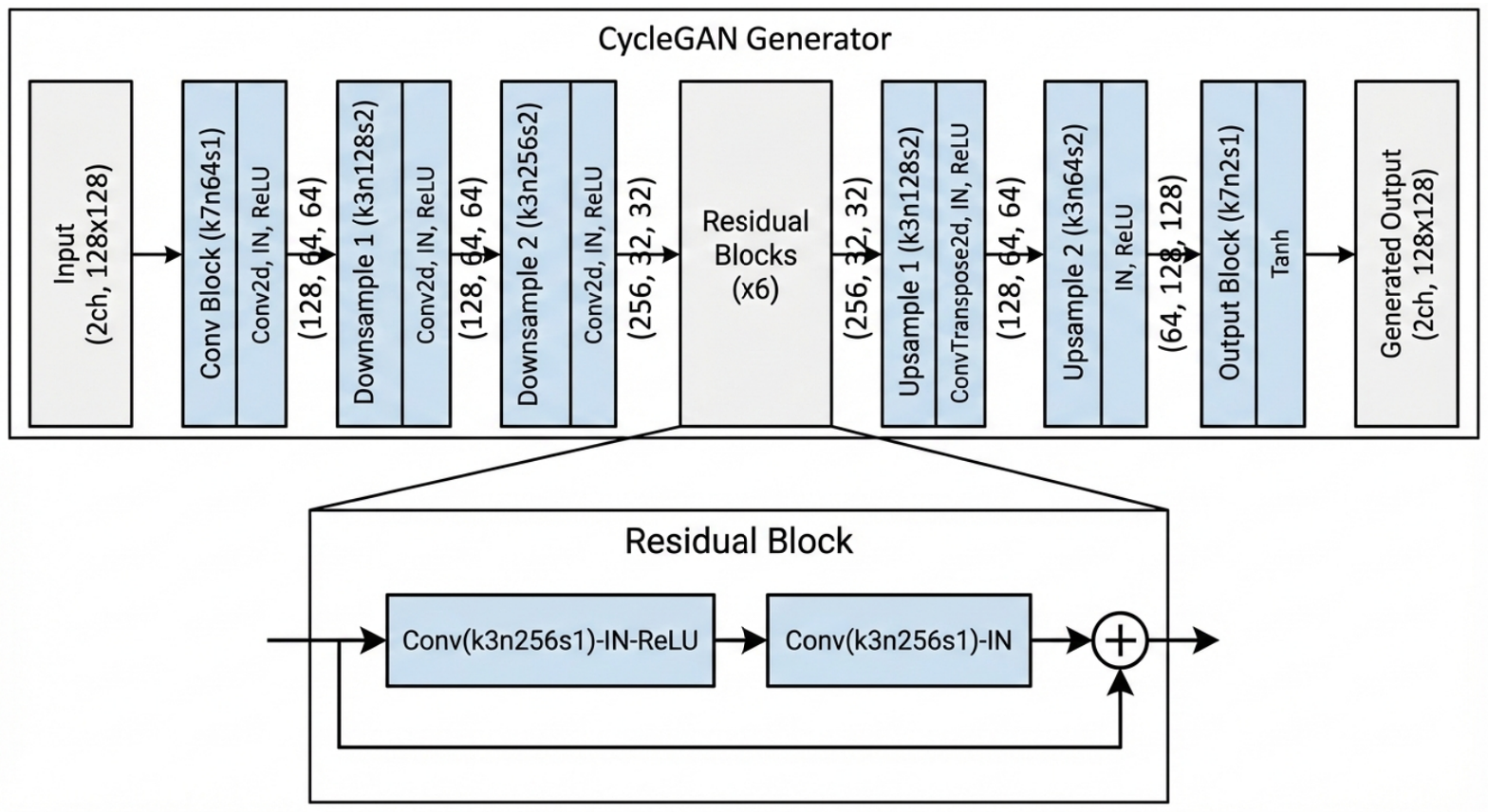} 
    \caption{\texttt{{Cycle-GAN}} Generator architecture. (inset) a zoom-in showing the layers composing each Residual Block. }\label{fig:generat}
    \end{figure*}

The generator (Fig.\ref{fig:generat})   employs  a {ResNet-based Encoder-Decoder} structure. The Encoder stage is made of 3 blocks of 2D convolutional layers, with a ReLU activation function,  aimed at  downsampling the image to extract "features" while discarding spatial redundancy. At the end of the Encoder stage, there are 6 Residual blocks with \emph{skip-connections }, so that the input in each block is added to the output  and  with a ReLU activation function. Finally, the Decoder reconstructs  processed features back into an image, via 3 consecutive transposed convolutional layers (or Deconvolutions). The output layer maps  the features back to  the  output channels, followed by a $\tanh$ activation function to force the values to range into the $[-1, 1]$ range.
Moreover,  we employ \emph{Instance Normalization} that  normalizes each image independently enhancing the absolute contrast of each  specific map. 

\begin{figure*}
\centering
    \includegraphics[width=1.5\columnwidth, angle=270]{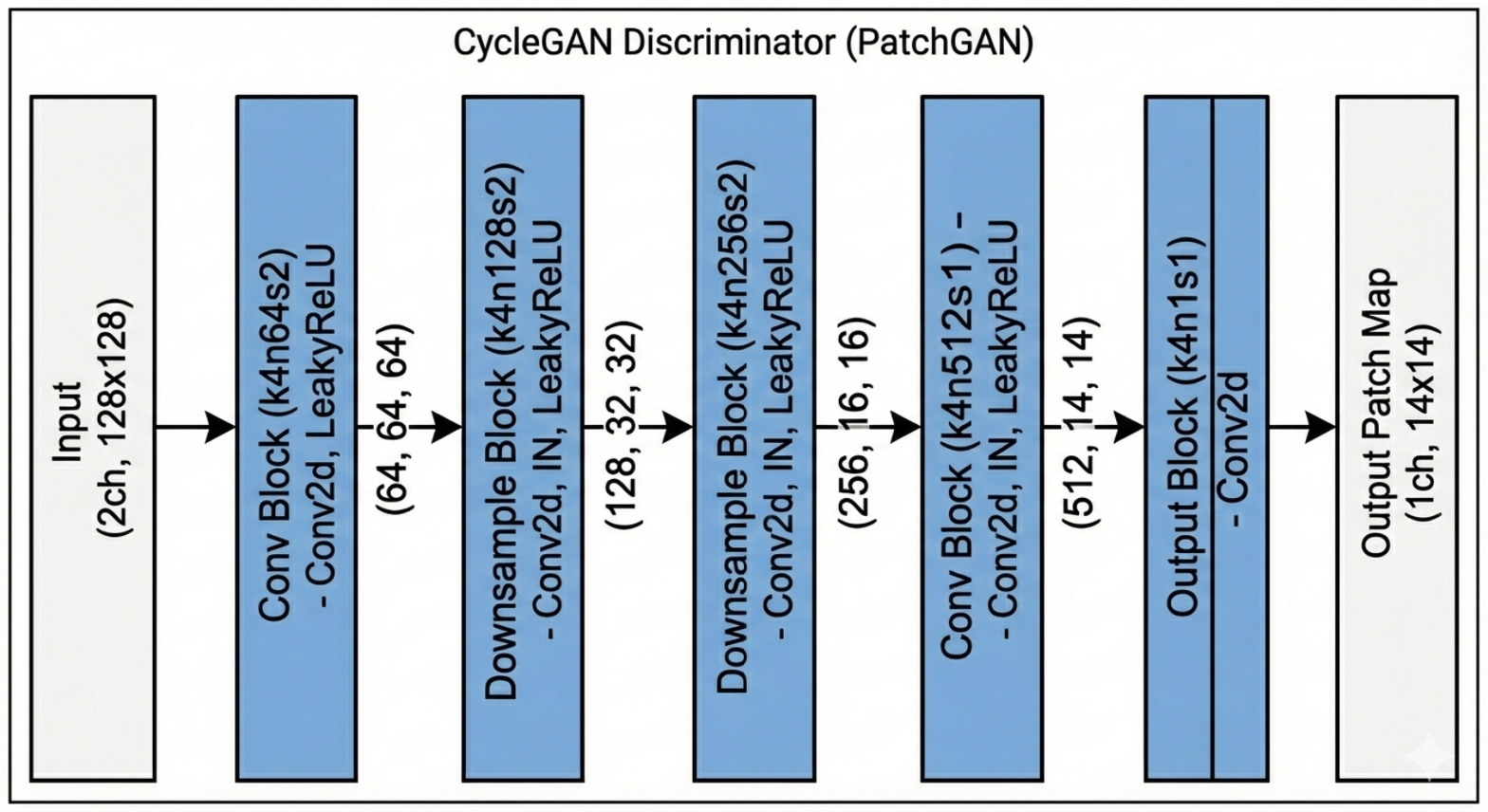}
    \caption{\texttt{Cycle-GAN} Discriminator architecture.  }\label{fig:discrim}
\end{figure*}

We choose the \emph{PatchGAN}  architecture  (Fig.\ref{fig:discrim}) for the discriminators, specifically  used in \texttt{Cycle-GAN}   (and Pix2Pix architectures \citep{isola2018}). A PatchGAN outputs a  grid (in our case $14\times 14$ ) of probabilities representing how much is likely to look  real a specific  patch in the input image. It is a fully convolutional network, and forces the Generator to produce high-frequency details, which are frequently lost in the context of image generation.  We employ LeakyReLU activation  function adjusted to the slope of $0.2$.


\begin{figure*}
\centering
     \includegraphics[width=2\columnwidth, trim=0 1cm 0 0 ,clip=true ]{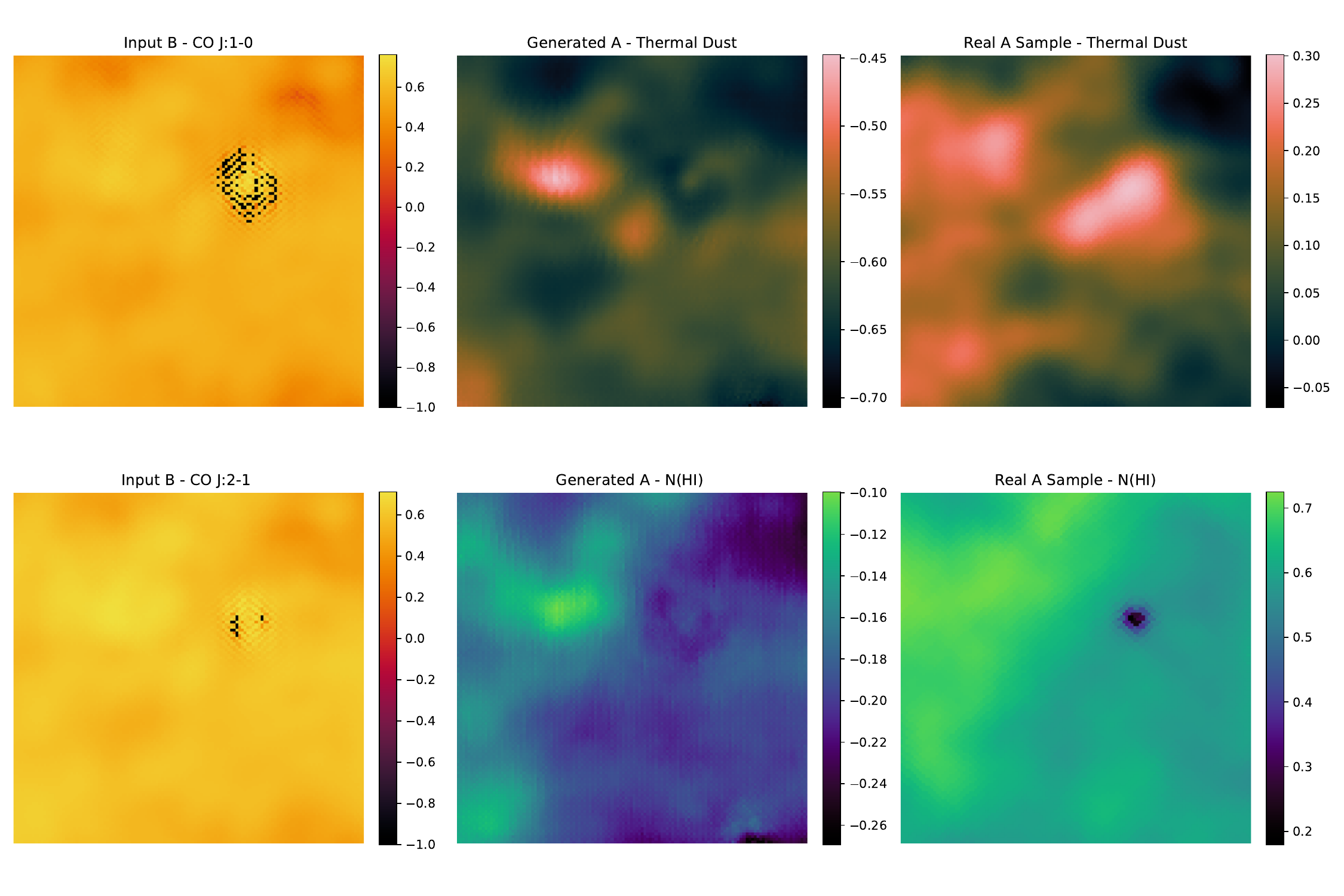}
          \includegraphics[width=2\columnwidth, trim=0 1cm 0 0 ,clip=true]{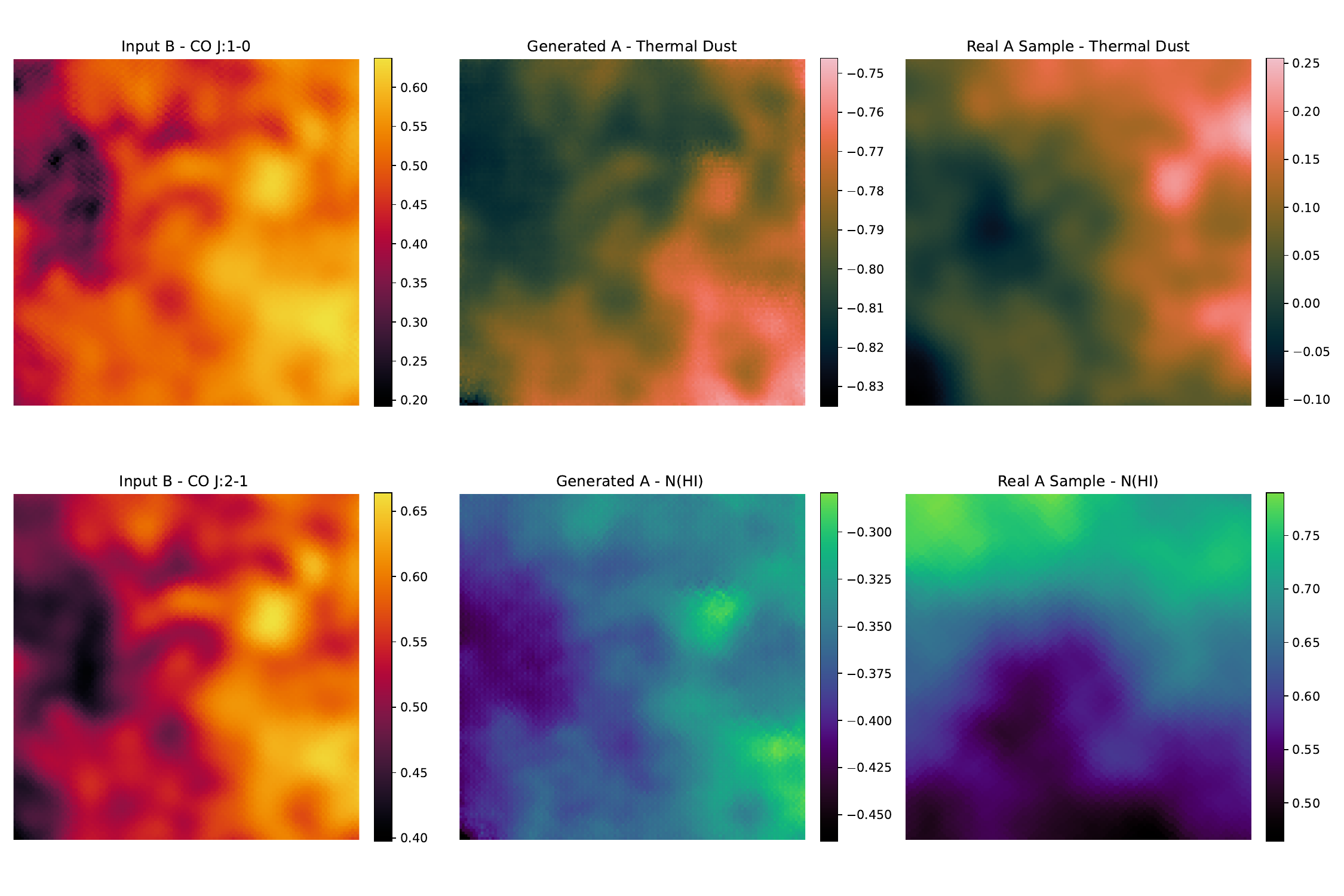}
\caption{Performance of the $G_{BA} $ generator, in this case the input features are the CO 1-0 and 2-1 maps (left column),  the generated and the output targets   of  thermal dust and $N(HI)  $ emissions, respectively in middle and right column.  }\label{fig:g_ba}
\end{figure*}

\bibliographystyle{elsarticle-harv} 
\providecommand{\sorthelp}[1]{}
\bibliography{example,refsPlanck,refsADS,refs}






\end{document}